\documentclass[sigconf]{acmart}

\usepackage{natbib}
\usepackage{amsmath,amsfonts}
\usepackage{algorithmic}
\usepackage{graphicx}
\usepackage{textcomp}
\usepackage{makecell}
\usepackage{enumitem}

\usepackage{url}

\usepackage{algorithmic}
\usepackage{graphicx}
\usepackage{textcomp}
\usepackage{hyperref}
\usepackage{listings}
\usepackage{cleveref}
\usepackage[T1]{fontenc}
\usepackage{float}
\usepackage{tabularx}
\usepackage{booktabs}
\usepackage{array,graphicx}
\usepackage{booktabs}
\usepackage{multirow}
\usepackage{graphicx}
\usepackage{siunitx}
\usepackage{nicematrix}
\usepackage{adjustbox}
\usepackage{enumitem}
\usepackage{pifont} 
\usepackage{tcolorbox}
\usepackage{comment}

\AtBeginDocument{%
  }
\setcopyright{acmlicensed}
\copyrightyear{2025}
\acmYear{2025}
\acmDOI{XXXXXXX.XXXXXXX}

\acmConference[EASE 2025]{The 28th International Conference on Evaluation and Assessment in Software Engineering}
\acmISBN{978-1-4503-XXXX-X/25/05}

\begin{document}



\title{Towards Automated Detection of Inline Code Comment Smells}

\author{Ipek Oztas}
\affiliation{%
  \department{Department of Computer Engineering}
  \institution{Bilkent University}
  \city{Ankara}
  \country{Turkey}}
\email{ipek.oztas@bilkent.edu.tr}

\author{U. Boran Torun}
\affiliation{%
  \department{Department of Computer Engineering}
  \institution{Bilkent University}
  \city{Ankara}
  \country{Turkey}}
\email{boran.torun@bilkent.edu.tr}

\author{Eray Tüzün}
\affiliation{%
  \department{Department of Computer Engineering}
  \institution{Bilkent University}
  \city{Ankara}
  \country{Turkey}}
\email{eraytuzun@cs.bilkent.edu.tr}

\renewcommand{\shortauthors}{Oztas, Torun and Tuzun}
\begin{abstract}
  \textbf{Background:} Code comments are important in software development because they directly influence software maintainability and overall quality. Bad practices of code comments lead to \textit{code comment smells}, negatively impacting software maintenance. Recent research has been conducted on classifying inline code comment smells, yet automatically detecting these still remains a challenge. \\
  \textbf{Objective: }We aim to automatically detect and classify inline code comment smells through machine learning (ML) models and a large language model (LLM) to determine how accurately each smell type can be detected. \\
  \textbf{Method: } We enhanced a previously labeled dataset, where comments are labeled according to a determined taxonomy, by augmenting it with additional code segments and their associated comments. GPT-4, a large language model, was used to classify code comment smells on both the original and augmented datasets to evaluate its performance. In parallel, we trained and tested seven different machine learning algorithms on the augmented dataset to compare their classification performance against GPT-4.\\
  \textbf{Results: } 
  %We used five measure metrics: Precision, Recall, F1-Score, Matthew’s Correlation Coefficient (MCC), and overall Model Accuracy to understand how well the models classify code comment smells. 
  The performance of models—particularly Random Forest, which achieved an overall accuracy of 69\%, along with Gradient Boosting and Logistic Regression, each achieving 66\% and 65\%, respectively —establishes a solid baseline for future research in this domain. The Random Forest model outperformed all other ML models, by achieving the highest Matthew’s Correlation Coefficient (MCC) score of 0.44. The augmented dataset improved the overall classification accuracy of the GPT-4 model's predictions from 34\% to 55\%. \\
  \textbf{Conclusion: } This study contributes to software maintainability by exploring the automatic detection and classification of inline code comment smells. We have made our augmented dataset and code artifacts available online, offering a valuable resource for developing automated comment smell detection tools.
\end{abstract}

\maketitle

\section{Introduction}\label{sec:section1}
%Introduction to the research problem or topic. 
%% context motivation code comment important reference to its importance
Code comments play a vital role in software development and directly affect software maintainability and code quality \cite{misra_reddy_chimalakonda_2020}. Since comments are mostly written in natural language, they serve as a linguistic bridge between the technical details of the code and human understanding, providing valuable insights to developers. A recent study has shown that code comments are correlated to solving issues in a software project \cite{misra_reddy_chimalakonda_2020}. Their study reveals issues are resolved faster in repositories with a higher percentage of relevant comments, suggesting a moderate positive correlation between relevant comments and issue resolution efficiency. Despite their importance, bad practices in code comments might raise issues as well. Yang et al. \cite{yang_jacky} argue that inconsistency between code and comments can confuse developers and lead to bugs. Within this context, we refer to ``comment smells'' as comments that can potentially lower software quality or do not aid developers in understanding the code. 

Notably, comments can be categorized as documentation comments and inline (implementation) comments \cite{nielebock_krolikowski_krüger_leich_ortmeier_2018}. In this study, we focus on inline comments, which are written within code blocks to clarify specific segments of the code, offering immediate explanations. A recent study by Jabrayilzade et al. \cite{elgun2024} has undertaken the classification of inline code comment smells and prepared a dataset using comments from four open-source Java projects and four open-source Python projects. According to their multi-vocal literature review, the taxonomy includes 11 types of inline code comment smells: \textit{misleading, obvious, no comment on non-obvious code, commented-out code, irrelevant, task, too much information, attribution, beautification, non-local} and \textit{vague}. 

Detecting these smells is essential for ensuring high software quality, and automation is instrumental in this process \cite{BLASI2021111069, Rabbi, LIU, Lin, tcomment}. Rani et al. \cite{rani_blasi_stulova_panichella_gorla_nierstrasz_2022} highlighted the importance of developing tools that can handle multiple programming languages to better assess and improve comment quality. Automated tools for detecting bad practices enhance the efficiency and accuracy of identifying undesirable patterns, such as code comment smells, which might otherwise be overlooked due to human error or oversight. 

Existing code comment smell detection approaches \cite{Lin, LIU, Rabbi, icomment, tcomment, wang_he_pal_marinov_zhou_2023} primarily focus on identifying a single type of comment smell: inconsistencies, in our terms misleading comments. To address a broader range of code comment smell types, we decided to work with the dataset labeled by Jabrayilzade et al. \cite{elgun2024}, which includes 2,448 labeled inline code comments. Since \textit{no comment on non-obvious code} and \textit{attribution} are not labeled in their dataset, we continued with the remaining nine categories. We conducted a manual review of 2,448 inline comments and examined their associated source codes. We extracted the relevant code segments for each comment to improve the dataset. We enhanced the pre-labeled dataset by adding the related code segments of the inline comments. 

%We made corrections and notified the authors when we discovered any possible mislabels.

Our study intends to use ML techniques to contribute to the automated detection of inline code comment smells. Combining data preprocessing, feature extraction, and oversampling approaches with the effectiveness of ML models like Naïve Bayes, Decision Trees, and Gradient Boosting, our approach aims to detect various types of code comment smells. We also utilize GPT-4, an LLM, to detect inline code comment smell types. In contrast to previous methods \cite{Lin, LIU, Rabbi, icomment, tcomment, wang_he_pal_marinov_zhou_2023, louis_dash_barr_sutton_2018} focusing on certain categories referred to as inconsistent or sub-optimal, our models are trained on an extensive dataset that includes nine different types of code comment smells.

%Problem statement or research question.\\
%Objectives and scope of the study.\\
%Significance and relevance of the research.\\
%explain what you are proposing and why it is interesting to write about. \\

We propose the following research questions to validate the effectiveness of our proposed solutions for detecting inline code comment smells:

%automated detection
\textbf{RQ1. Can we automatically detect inline code comment smells?}

%The automation of code comment smell detection is motivated by software projects' increasing size and complexity. Automating this process allows developers to focus on more substantial tasks while ensuring code comments follow best practices. This research question investigates the impact of various ML algorithms and the GPT-4 model in automating the identification of inline code comment smells. It evaluates these algorithms' performance in terms of several metrics in order to provide guidance for the creation of an application that can be integrated into software development environments.

%How do different machine learning algorithms perform for automated code comment smell detection?
%{RQ. How can we customize ML models to perform better detecting specific types of code comment smells?
%\textbf{RQ. Are there any features or heuristics that can improve code comment smell detection?}
%\textbf{RQ. How well do the proposed models perform with larger datasets?}
%\textbf{RQ. How does the diversity of the dataset affect the model's performance in terms of accuracy, precision, and recall?}

%integration
\textbf{RQ2. Does dataset augmentation, which provides additional information, affect the prediction of the GPT-4 model?}
 
%Dataset augmentation refers to adding related code segments to inline comment entries in the dataset of Jabrayilzade et al. \cite{elgun2024}, manually checking each ground truth label, and correcting if necessary. This research question investigates the impact of dataset augmentation on the prediction of the GPT-4 model for inline code comment smells. By augmenting the dataset, we aim to assess changes in the performance of GPT-4 model predictions using metrics including accuracy, precision, recall, F1-score, and MCC.

% Contributions of the paper
Contributions of this paper include:
\begin{itemize}
    \item [1] We augment a previously labeled inline comment smells dataset with the associated code segments, working on a wider variety of code comment smell types compared to the existing literature.
    
    \item [2] We benefit from seven ML classifiers for the automated detection of inline code comment smells, addressing five out of nine smell types, which were identified after removing the minority classes.

    \item [3] We share our replication package, including our code artifacts to train ML classifiers, prompts for the GPT-4 model, and augmented inline code smells dataset, including 2,211 code segment-inline comment pairs (after the removal of the duplicates where the comment, and label are identical), which can be used for future work.\footnote{ \url{https://figshare.com/s/49ae815ab21d916726ab}}
\end{itemize}

%Outline of the paper's structure.
%The rest of the paper is organized as follows:
%\hyperref[sec:section2]{Section 2} presents the necessary explanation of inline code comment smells with the related taxonomy. Related work is also presented in this section. We explain the classification of inline code comment smells and present the prior research highlighting the notable work. \hyperref[sec:section3]{Section 3} details the methodology and steps followed for answering the research questions. In \hyperref[sec:section4]{Section 4}, we report the results obtained from our study and discuss their outcomes in \hyperref[sec:section5]{Section 5}. Then, \hyperref[sec:section6]{Section 6} discusses the possible threats to the validity of our study. Finally, \hyperref[sec:section7]{Section 7} serves as the conclusion, highlighting the key insights from our study and outlining promising implications for future work.

%Detailed background information related to the research topic. \\
%Info about comment smell\\
%Historical context (if applicable).\\
%Key concepts and terminology.\\

\section{Background and Related Work}\label{sec:section2}
 We begin with the classification of inline code comment smells, focusing on Jabrayilzade et al.'s \cite{cohen_1960} taxonomy of 11 types. Since code comment inconsistencies are the most common in the literature, we then explore methods for detecting them using natural language processing, machine learning, and program analysis. Finally, we discuss automated approaches for maintaining comment consistency during code changes.

%In this section, we review the existing literature on detecting inconsistencies in code comments, setting the stage for our research goals. Section \ref{sec:section2.1} outlines the taxonomy adopted from a recent study.

%Overview of existing literature and research related to the topic. \\
%Comparison of previous work and your research objectives.\\
%Identification of gaps in the literature.\\
%Explain what other authors have done how that work relates to your own contributions.\\
%smell types subsection\\

\subsection{Classification of Inline Code Comment Smells}\label{sec:section2.1}

Jabrayilzade et al. \cite{elgun2024} propose a taxonomy for inline code comment smells, including 11 smell types based on a literature review, which included 55 sources from gray literature and 29 from white literature. The taxonomy is presented in the replication package.\footnote{\url{https://figshare.com/s/49ae815ab21d916726ab?file=49678578}} They manually annotated 2,448 inline code comments from four Java and four Python open-source projects. They further investigate the importance of inline code comment smells by surveying software practitioners and creating pull requests to remove comment smells.

Given that prior studies \cite{wang_he_pal_marinov_zhou_2023, icomment} concentrated on Java comments and addressed only a single type of smell, we decided to adopt Jabrayilzade et al.'s \cite{elgun2024} taxonomy as the baseline and prioritize detecting inline comment smell types included in the dataset.

%Corazza et al. \cite{corazza2018coherence} provided a manually curated dataset highlighting coherence issues between Java method comments and implementations, emphasizing the significance of coherence as a quality indicator.

%Code comment smell occurs when comments fail to accurately describe or reflect the underlying code, leading to misunderstandings during code maintenance and review. This inconsistency can arise due to outdated comments not being updated after code changes or comments that do not adequately explain the logic of a code segment, making it crucial to detect and address them to improve code comprehension. Previous research on inconsistent code comment detection is detailed in Section \ref{sec:section2.2}, and automated approaches for maintaining comment consistency following code changes are discussed in Section \ref{sec:section2.3}.

\subsection{Code Comment Inconsistency Detection}\label{sec:section2.2}

Tan et al. \cite{icomment} introduced \textit{icomment}, which combines Natural Language Processing (NLP) techniques, ML, and program analysis techniques to analyze comments and detect inconsistencies between source code and comments. \textit{icomment} extracts implicit rules from comments and identifies discrepancies between these rules and the actual code. Their approach enhanced the identification of \textit{misleading} comment smell type, contributing to more efficient code-comment analysis.

Tan et al. \cite{tcomment} also introduced \textit{@tComment}, which focuses on detecting inconsistencies between Javadoc comments and Java method bodies. They utilized information extraction from Javadoc comments to infer properties and generate random tests to check for inconsistencies. This approach led to discovering inconsistencies across several open-source projects, emphasizing the importance of maintaining consistency between comments and code.

Rabbi and Siddik \cite{Rabbi} proposed a novel solution to detect code-comment inconsistencies by utilizing a Siamese recurrent network, a combination of two identical neural networks trained simultaneously. Their approach involves processing code and comments through separate Recurrent Neural Network-Long Short-Term Memory (RNN-LSTM) models to convert them into vector representations. The model then measures the similarity between the vectors, identifying inconsistencies based on discrepancies in sequence order and word tokens. This approach detected inconsistencies in a dataset of 2,881 code-comment pairs.

Blasi et al. \cite{BLASI2021111069} introduced a tool called \textit{RepliComment}, a tool designed for detecting duplicate comments and pointing out the comments that require correction. Their approach consists of analyzing code comments to detect clones that arise from copy-paste errors or poor documentation practices. \textit{RepliComment} leverages natural language processing to differentiate between comment clones and those that require correction. Their tool detected over 11,000 instances of comment clones across ten well-known open-source Java projects.

Wang et al. \cite{wang_he_pal_marinov_zhou_2023} referred to low-quality code comments as ``Suboptimal Comments'', which significantly impact code comprehension and maintenance. They focused on Independent Comment Changes (ICCs) in Java, defining ICCs as modifications to comments that occur without accompanying code changes. The study emphasized the role of tools in addressing suboptimal comments. It was found that while tools like Javadoc are commonly integrated into projects and can reduce the proportion of ICCs, indicating improved comment quality, other tools like Checkstyle did not significantly reduce ICCs.

Louis et al. \cite{louis_dash_barr_sutton_2018} suggested that comments in software that are easily inferable from the code are generally ``redundant''. They introduced the concept of ``comment entailment'', where a comment is considered entailed if its content can be fully inferred from the associated code. They developed a tool named Craic, which is based on sequence-to-sequence (seq2seq) deep learning models \cite{sutskever2014sequencesequencelearningneural}. These models are trained to generate comments based on code snippets. The aim was that if a model can easily generate a comment from the code, then the comment is likely redundant. However, their study only evaluated comments in Java projects and failed to detect a broader range of comment smells.

Igbomezie et al. \cite{igbomezie2024simplicity} proposed Co3D, a neural approach utilizing word embeddings and LSTM to detect coherence between code comments and method implementations, outperforming traditional models.

While most studies have primarily focused on detecting misleading comment smells and have generally been limited to Java, our study broadens the scope by addressing a wider range of comment smell types across both Java and Python.

\subsection{Automated Comment Consistency Maintenance Approaches}\label{sec:section2.3}
When there is a code change, developers should also update the related comment segment to prevent inconsistencies. Just-in-time comment updating is critical for maintaining accurate documentation with evolving code. Several studies have explored automated techniques for ensuring comment consistency, highlighting the importance of keeping documentation in sync with code changes.

Liu et al. \cite{LIU} introduced CUP (Comment Updater), an automated just-in-time comment update technique. CUP leverages a neural sequence-to-sequence model to learn patterns from existing code-comment co-changes. It then automatically updates comments based on code changes and the previous comment. The tool effectively detects outdated comments, providing suggestions for updates during code changes. In their evaluation on a dataset of over 108,000 code-comment pairs, CUP outperformed rule-based and retrieval-based baselines, and its automatically updated comments matched developer-written comments in 16.7\% of cases.

Lin et al. \cite{Lin} proposed HEBCUP, a heuristic-based comment updater, which outperforms CUP in various metrics. They suggested heuristics focusing on the modified code and performing token-level comment updates.

Yang et al. \cite{yang_jacky} also contributed to this field by proposing a composite approach named CBS (Classifying Before Synchronizing). CBS seeks to further enhance the code-comment synchronization performance by combining the benefits of CUP and HebCUP with the aid of inferred categories of Code-Comment Inconsistent (CCI) samples.

Panthaplackel et al. \cite{panthaplackel_nie_milos_gligoric_junyi_jessy_li_mooney_2020} aimed to automate the updating of natural language comments in software, ensuring they stay relevant with changes made to the code. They employed a sequence-to-sequence model with attention mechanisms that learn to generate edit actions from code modifications, thereby aligning comments with code updates. However, the study lacks generalization across different programming languages.

\section{Methodology}\label{sec:section3}

%lucidchart
\begin{figure*}
    \centering
    \includegraphics[width= \textwidth]{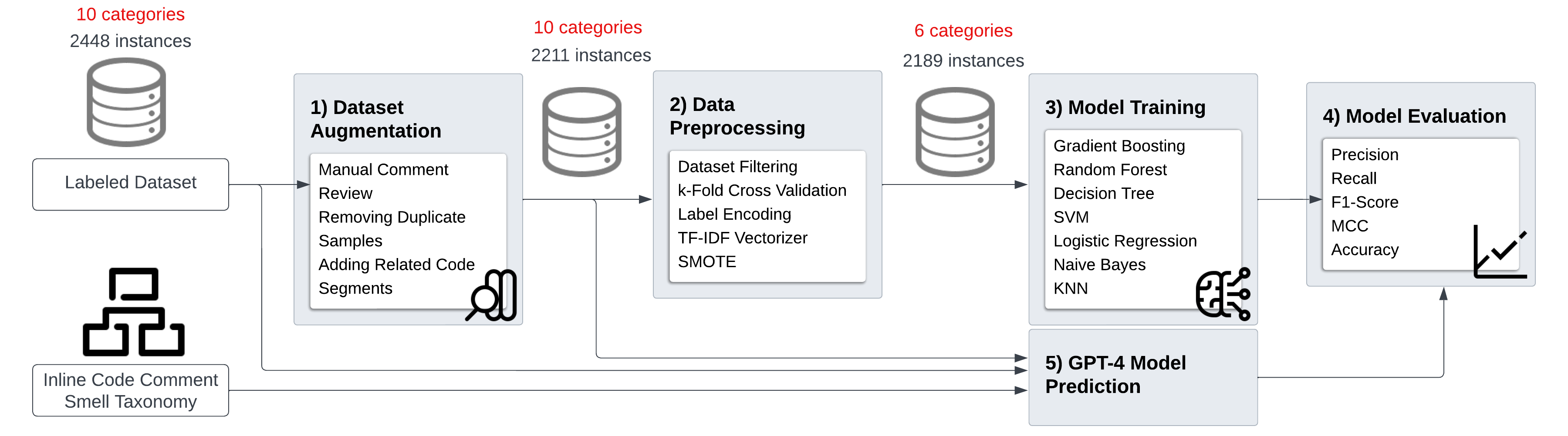}
    \caption{Proposed research approach}
    \Description{}
    \label{fig:methodology}
\end{figure*}

%Explanation of data collection methods.
%Overview of sampling procedures (if applicable).
%Presentation of data analysis techniques.
%Ethical considerations (if applicable).
%Forward and backward snowballing for MLR

%Description of the research design. 

%Our study uses the taxonomy of inline code comment smells by Jabrayilzade et al. \cite{elgun2024} as a baseline for classification and employs their dataset to further analyze with ML models. 
%The dataset provided by Jabrayilzade et al. includes 2,448 pairs of comments and their associated smell types. Additionally, we enriched this dataset by adding code sections to observe how including code affects prediction improvements using the GPT-4 model. We trained seven ML classifiers to detect inline code comment smells using the augmented dataset. 

A schematic representation of the proposed workflow for detecting code comment smells is shown in Figure \ref{fig:methodology}. In this figure, the red numbers represent the number of categories, and the black numbers indicate the number of instances at each step of the process. In the following subsections, we provide details about our methodology.

\subsection{Dataset Description}\label{sec:section3.1}
The dataset that we use, which is proposed by Jabrayilzade et al. \cite{elgun2024} consists of 2,448 manually labeled inline comments extracted from eight open-source projects, equally divided between four Java projects (Anki-Android, Jitsi, Moshi, and Light-4j) and four Python projects (Requests, Scrapy, Kivy, and Scikit-learn). The original dataset includes source code, comment smells, and their associated label, but the scope of the comments is not defined. Scope in this context refers to the specific boundaries or coverage of the comments in relation to the code. The granularity of the instances in the dataset is at the file level, which means that inline comments were sampled along with their surrounding code to preserve the context for accurate labeling. The average length of the comments in the dataset is 61.98 characters. However, there is a class imbalance in the dataset. The distribution of instances across the defined categories is shown in the replication package.\footnote{\url{https://figshare.com/s/49ae815ab21d916726ab?file=51674633}}

\subsection{Dataset Augmentation}\label{sec:section3.2}

We extend our methodology to include the augmentation of the dataset by incorporating related code segments with the most accurate scope, aiming to enhance the classification performance of the GPT-4 model. The code segment selected for association with a comment is determined based on its relevance and accuracy in capturing the intended scope of the comment. 

\subsubsection{Manual Comment Review: }As the first step, we reviewed all comments in the dataset individually. During this review process, the content of each comment is comprehended and interpreted.

\subsubsection{Removing Duplicate Samples:} We removed duplicates by implementing an automated function that checks for identical (comment string, label string) pairs and discards them, ensuring no manual intervention was required. This process effectively eliminated duplicates while preserving data integrity. The size of the dataset decreased from 2,448 to 2,211.

\subsubsection{Adding Related Code Segments: } As a second step, we determine the relevant code segments that reflect the purpose of each comment and capture its correct scope. Then, we manually inserted the related code segments for 2,211 comments. This process involves analyzing the semantic context of the comment and identifying the code segment that best summarizes the meaning of the comment. These code segments help us better understand the context addressed by the comment. 

We followed a manual approach to determine the scope of associated code, as manual annotation typically results in higher accuracy compared to automated methods. This decision aligns with the findings of Chen et al. \cite{chen_huang_liu_chen_zhou_luo_2019}, which highlights the challenges and importance of accurately determining comment scopes to improve the performance of ML models. Chen et al. \cite{chen_huang_liu_chen_zhou_luo_2019} utilize a ML model, Random Forest, with the features of code snippets and comments to detect comment scopes automatically, which achieves an accuracy of 81.15\%. They applied heuristic rules to divide the code into code-comment pairs to form their dataset.

However, including the study of Chen et al. \cite{chen_huang_liu_chen_zhou_luo_2019}, no publicly available code for detecting the scope of code comments exists, necessitating our manual approach. In our study, we also employed a set of heuristics to associate code snippets with comments based on the type of code comment smell. These heuristics were designed to ensure that the code snippet accurately reflects the context of the comment, thereby aiming to enhance the performance of our models. The heuristics categorized by the type of comment smell are as follows:

\begin{enumerate}
    \item \textbf{Beautification, Commented-out Code, and Task:} For these categories, the content of the code snippet is irrelevant to the nature of the comment. Therefore, we did not insert any associated code snippet. Instead, we augmented these comments with "NA" (Not Applicable) to indicate that no code context is necessary.
    \item \textbf{Obvious, Vague, Misleading, Too much information, Non-local information, Irrelevant, and Not a smell: }The primary goal for these categories was to provide the model with the most relevant part of the code that corresponds to the comment. We opted for a larger scope to provide the model with more comprehensive information. Heuristics categorized by the code scope are explained as follows:
    \begin{itemize}[nosep, leftmargin=*]
        \item Single Line Scope: If the comment applies to a single line of code, only that line was taken as the snippet. When the comment is immediately followed or preceded by code, the corresponding snippet is taken as the line directly above or below the comment, depending on its position.
        \item Block-Level Scope: If the comment pertains to a specific block of code (e.g., within a method or loop), the entire block was taken as the corresponding code snippet.
        \item Other Cases: In situations where the scope of the comment was more ambiguous, the final decision on the code snippet was made collaboratively by the authors, ensuring that the selected code best reflects the intent of the comment.
    \end{itemize}

\end{enumerate}

%\textbf{Mislabel Corrections: }Comments identified as mislabeled during the manual %review process are corrected. This is important for improving the accuracy of the %dataset and contributes to better algorithm performance. 

%Corrections included are deletion of a comment because it was a code segment %rather than a comment, change of the labeling from \textit{obvious} to %\textit{vague}, and labeling of a comment as \textit{vague}, which was previously %labeled as \textit{not a smell}.

Based on these heuristics, the first author, with five years of experience in Java and Python, prepared an augmented dataset, which was then reviewed by the second author, who also has five years of experience in Java and Python. The second author had full access to the complete dataset, enabling a thorough and contextual validation of the selected code segments compared to relying solely on the extracted snippets of the first author. The third author, with 15+ years of experience in both Java and Python, later participated in resolving disagreements. Here, disagreements are defined as cases where annotators did not select the same lines of code for a given comment. For example, if one annotator chose two lines of code for a comment and another chose five lines for the same comment, it was counted as a disagreement. Out of the 2,211 comments, there were disagreements in 250 comments, corresponding to an 88.69\% agreement rate. To address these disagreements, a session was held with the third author to review and resolve conflicts. The third author’s opinion helped annotators converge on the same set of lines for each comment. These issues were subsequently addressed to ensure the dataset's accuracy and consistency. In addition, the calculation of Inter-Rater Reliability (IRR) was not feasible due to inconsistencies in selecting the code field. Calculating the IRR score in cases where the output space (i.e., corresponding code segments) can take any value is challenging, especially since traditional IRR measures like Cohen's Kappa \cite{cohen_1960} or Krippendorff’s Alpha \cite{krippendorff_2011} assume a fixed set of possible outcomes. The problem arises from the same reasons that caused disagreements, such as annotators selecting different lines of code for the same comment. Instead, a thorough quality control check, as explained above, was conducted to resolve these issues and ensure consistency in the dataset.

%These findings indicate areas where the dataset needed refinement to ensure accuracy and consistency. 

After the addition of related code segments, the final version of the augmented dataset is created. Through this augmentation process, we expect to improve the accuracy and effectiveness of the GPT-4 model in identifying and classifying various types of inline code comment smells. We retained the minority classes because GPT-4 is a pre-trained model, allowing it to handle a broader range of data without the same risk of overfitting as training ML models.

\subsection{Data Preprocessing}\label{sec:section3.3}

Data preprocessing is conducted to prepare the original dataset from Jabrayilzade et al.'s study \cite{elgun2024} only for training the ML models.

\textbf{Removing Minority Classes:} Minority classes, defined as categories with fewer than 30 instances, were discarded. This step is crucial because ML models require a sufficient number of examples from each class to learn effectively. This step is essential to prevent the model from being biased towards the majority classes. Although SMOTE oversampling \cite{chawla_bowyer_hall_kegelmeyer_2002} is employed to address class imbalance, classes with fewer than 30 instances may not be sufficiently representative and would introduce inefficiencies and potential noise during the oversampling process. Fitzner and Heckinger state that although there is not a definitive threshold for what constitutes a small sample size, statistical challenges can occur when the sample consists of fewer than 30 subjects \cite{fitzner_heckinger_2010}. A sample of this size often lacks diversity and fails to adequately represent a normal distribution. Therefore, retaining these classes would be impractical for achieving robust model performance. Additionally, the number of smell types was reduced from nine to five. Consequently, the ML models only classify the input as one of the following categories: \textit{obvious, task, beautification, vague, commented-out code} or \textit{not a smell}. Four of the smell types (\textit{non-local information, misleading, irrelevant, and attribution}) were removed. The number of instances decreased from 2,211 to 2,189. 

Notably, most instances, accounting for 57.5\%, are labeled as \textit{not a smell}. Among the various smell types, \textit{obvious} is the most prevalent, constituting 30.8\% of the dataset. The remaining smell types, namely \textit{task, vague, beautification}, and \textit{commented-out code}, represent proportions of 5.5\%, 2.3\%, 2.2\%, and 1.6\%, respectively.

\textbf{Label Encoding:} The textual class labels are converted to numerical values using the scikit-learn LabelEncoder.\footnote{ \url{https://github.com/scikit-learn/scikit-learn/blob/main/sklearn/preprocessing/_encoders.py}} First, the training labels are used to fit the label encoder. To ensure the test data does not affect the training process, the training labels are encoded using the fit transform method, and the test labels are encoded using the transform method.

\subsection{Feature Extraction}\label{sec:section3.4}

In this study, we employed Term Frequency-Inverse Document Frequency (TF-IDF) \cite{spärck_jones_2004} to represent the textual content of inline code comments for machine learning models. The corpus was preprocessed by removing stop words, converting all text to lowercase, and applying tokenization. Terms were extracted based on unigram features, and document frequency calculations were performed by counting occurrences throughout the data set. TF-IDF was selected due to its interpretability and effectiveness in highlighting important terms in smaller, domain-specific datasets. While recent advanced approaches, such as embeddings (e.g.,  word2vec \cite{mikolov2013efficient}, GloVe \cite{pennington2014glove}, and transformer-based models \cite{vaswani2017attention}), can capture contextual meaning, they require significantly larger datasets to generalize well. Given the size and nature of our dataset, TF-IDF provides an effective representation.

% \textbf{Heuristics}: Heuristics were applied as additional guidelines to refine the feature extraction process. These rules or strategies aim to capture specific patterns, structures, or characteristics within the code comments of smell types \textit{beautification} and \textit{task}, allowing for a more context-aware representation of the numerical features. For the \textit{beautification} comment smell type, it was observed that most of the occurrences included large tokens with characters such as ``----'' and ``****''. These tokens, as well as the length and the number of tokens, were taken into consideration while constructing the heuristics. For the \textit{task} comment smell type, the comments containing at least one of 'todo', 'fixme', 'github', 'tracked' words were classified as \textit{task} comment smell. This integration of heuristics proposes a simpler and more efficient way of detecting \textit{task} and \textit{beautification} comment smell types.

\subsection{ML Model Training}\label{sec:section3.5}

For the inline comment smell classification task, we trained traditional ML classifiers to establish a baseline, as there is no prior work on multiclass classification for code comment smells. Unlike LLMs, ML models cannot fully incorporate the taxonomy, and we cannot observe any improvement. So, we used ML classifiers primarily to benchmark the augmented dataset only. To ensure a robust performance evaluation, we trained and evaluated seven ML models using a stratified 10-fold cross-validation approach. We chose stratified 10-fold cross-validation to ensure that each fold maintains the original distribution of classes, which is especially important given the class imbalance in our dataset. This approach provides a reliable estimate of model performance while minimizing the variance that may result from a single train-test split.

The \textbf{Naive Bayes} algorithm, specifically the multinomial variant, was chosen for its efficiency in text-based classification tasks and was trained with default parameters. Similarly, \textbf{Logistic Regression} was chosen for its simplicity in finding optimal decision boundaries based on comment features.

We included the \textbf{Random Forest} classifier, which aggregates decision trees to improve generalization, alongside a standalone \textbf{Decision Tree} model for interpretable decision-making.

The \textbf{Support Vector Machine (SVM)} model was trained to leverage its robust hyperplane separation in high-dimensional spaces.

\textbf{Gradient Boosting}, another ensemble method, iteratively corrects errors from previous trees, learning complex patterns.

Finally, the \textbf{K-Nearest Neighbor (KNN)} algorithm was employed for its approach of using the nearest neighbors for classification, and number of neighbors parameter was set to three.

Each model was trained using the following methodology:

For text vectorization, The TF-IDF vectorizer \cite{spärck_jones_2004} was applied after splitting the data into training and testing sets using a ratio of 80:20 to avoid data leakage.

To handle class imbalance during training, we used SMOTE \cite{chawla_bowyer_hall_kegelmeyer_2002} to generate synthetic samples for the minority classes, ensuring that the models were trained on a more balanced dataset. However, it is important to note that SMOTE was only applied to the training data. The original class distribution was retained for the testing data to ensure that the models were evaluated in realistic conditions, preserving the natural imbalance present during inference.

A stratified 10-fold cross-validation was performed, where each fold preserved the percentage of samples for each class. MCC was chosen as the primary evaluation metric due to its suitability for imbalanced datasets, alongside accuracy.

For each fold, the models were trained on the resampled training data and evaluated on the validation set. The mean MCC and accuracy scores across all folds were computed for comparison.

%\subsection{Evaluation of ML Models}\label{sec:section3.6}
After training, the models were evaluated on the test set, and metrics, including accuracy, precision, recall, F1 score, and MCC, were reported. The use of a pipeline ensured that the same preprocessing steps (e.g., vectorization and oversampling) were consistently applied across all models and cross-validation folds. The results were compared to determine the most effective model for the classification task.

\subsection{Evaluation of GPT-4}\label{sec:section3.6}
We explored the application of OpenAI GPT-4, a transformer-based large language model, in the detection of code comment smell. This model was selected due to its natural language understanding and code analysis capabilities \cite{openai2024gpt4technicalreport}. Our aim was to evaluate the accuracy of the model in identifying code comment smells according to the taxonomy that we provide to the model. 
As mentioned in Section \ref{sec:section1}, the adopted dataset includes nine comment smell types out of the 11 types proposed in the taxonomy \cite{elgun2024}.
The prompt design includes a predefined set of ten categories consisting of nine inline comment smell types and the \textit{not a smell} category. Each smell type is specified in the prompt, along with its description and example, as shown in the replication package.\footnote{\url{https://figshare.com/s/49ae815ab21d916726ab?file=49678578}}

The OpenAI API is invoked to generate responses based on provided prompts, and the results are compared against ground truth labels to assess the model's accuracy in identifying code comment smells. We also configured three parameters of the model, including \textit{temperature}, \textit{max tokens}, and \textit{top\_p}. Adjusting these values provided us control over the model according to our task. Lower \textit{temperature} values make the model's responses more deterministic and predictable \cite{vega_2023}, which is useful for classification purposes. We chose \textit{temperature} to be 0.2 since OpenAI documentation states that the use of \textit{temperature} between 0-0.2 for more focused and deterministic responses \cite{openai}. \textit{Max tokens} limits the number of tokens in the model’s response, which is determined using the tokenizer provided by OpenAI \cite{openai}. We selected a \textit{max tokens} value of 10 to ensure very concise responses, as the task focuses on the classification of code comment smells, which typically do not require lengthy outputs. The longest class label in our dataset, "commented out code," consists of five tokens according to the OpenAI tokenizer. To allow potential variations, we added a safety margin, resulting in a final \textit{max tokens} value of 10. \textit{Top\_p}, also referred to as ``nucleus sampling'' determines the number of possible words to consider \cite{vega_2023}. Since the prompt includes a predetermined taxonomy, we chose the \textit{top\_p} value as low as possible, 0.1.

We prompted the model, both the original dataset and augmented dataset (which includes the associated code segment of comment) introduced in Section \ref{sec:section3.1}, to compare the improvement of its detection performance. Thus, GPT-4 generates feature representations for both code and natural language, and we expect more accurate predictions when code and comments are present. For space restrictions,  we could not include examples of input prompts and outputs here. For examples of these, please refer to the replication package.\footnote{\url{https://figshare.com/s/49ae815ab21d916726ab}}

\section{Results}\label{sec:section4}

%This section will present the dataset distribution, as well as the performance of ML models and the GPT-4 model.

\subsection{Performance of ML Models}

We present results of ML models from the scikit-learn package for detecting comment smells in the augmented version of Jabrayilzade et al.'s dataset \cite{elgun2024}. In this part of the experiment, seven models are trained to detect code comment smells.

\begin{table*}[t]
\centering
\caption{Performance of the selected seven ML models}
\label{tab:metrics}
\resizebox{0.8\textwidth}{!}{
\begin{tabular}{@{}llllllll@{}}
\toprule
Model & Not a smell & Obvious & Task & Beautification & Vague & Commented-out code & Total \\ \midrule

Gradient Boosting & 
\begin{tabular}[c]{@{}l@{}}P: 0.71 \\ R: 0.83 \\ F1: \textbf{0.77}\end{tabular} & 
\begin{tabular}[c]{@{}l@{}}P: 0.56 \\ R: 0.48 \\ F1: 0.52\end{tabular} & 
\begin{tabular}[c]{@{}l@{}}P: 1.00 \\ R: 0.67 \\ F1: \textbf{0.80}\end{tabular} & 
\begin{tabular}[c]{@{}l@{}}P: 0.00 \\ R: 0.00 \\ F1: 0.00\end{tabular} & 
\begin{tabular}[c]{@{}l@{}}P: 0.50 \\ R: 0.50 \\ F1: 0.50\end{tabular} & 
\begin{tabular}[c]{@{}l@{}}P: 0.50 \\ R: 0.33 \\ F1: 0.40\end{tabular} & 
\begin{tabular}[c]{@{}l@{}}P: 0.67 \\ R: 0.67 \\ F1: 0.66\end{tabular} \\ \midrule

Random Forest & 
\begin{tabular}[c]{@{}l@{}}P: 0.71 \\ R: 0.85 \\ F1: \textbf{0.77}\end{tabular} & 
\begin{tabular}[c]{@{}l@{}}P: 0.64 \\ R: 0.45 \\ F1: 0.53\end{tabular} &  
\begin{tabular}[c]{@{}l@{}}P: 1.00 \\ R: 0.44 \\ F1: 0.62\end{tabular} &  
\begin{tabular}[c]{@{}l@{}}P: 0.40 \\ R: 1.00 \\ F1:\textbf{ 0.57}\end{tabular} &  
\begin{tabular}[c]{@{}l@{}}P: 0.67 \\ R: 0.50 \\ F1:\textbf{ 0.57}\end{tabular} &  
\begin{tabular}[c]{@{}l@{}}P: 1.00 \\ R: 0.67 \\ F1:\textbf{
0.80} \end{tabular} &  
\begin{tabular}[c]{@{}l@{}}P: \textbf{ 0.70} \\ R: \textbf{0.69} \\ F1: \textbf{0.67}\end{tabular} \\ \midrule

Decision Tree & 
\begin{tabular}[c]{@{}l@{}}P: 0.74 \\ R: 0.74 \\ F1: 0.74\end{tabular} &  
\begin{tabular}[c]{@{}l@{}}P: 0.57 \\ R: 0.55 \\ F1: 0.56\end{tabular} &  
\begin{tabular}[c]{@{}l@{}}P: 0.75 \\ R: 0.67 \\ F1: 0.71\end{tabular} &  
\begin{tabular}[c]{@{}l@{}}P: 0.40 \\ R: 1.00 \\ F1:\textbf{ 0.57}\end{tabular} &  
\begin{tabular}[c]{@{}l@{}}P: 0.67 \\ R: 0.50 \\ F1:\textbf{ 0.57}\end{tabular} &  
\begin{tabular}[c]{@{}l@{}}P: 1.00 \\ R: 0.67 \\ F1:\textbf{
0.80}\end{tabular} &  
\begin{tabular}[c]{@{}l@{}}P: 0.67 \\ R: 0.67 \\ F1: 0.67\end{tabular} \\ \midrule

SVM & 
\begin{tabular}[c]{@{}l@{}}P: 0.63 \\ R: 0.89 \\ F1: 0.74\end{tabular} &  
\begin{tabular}[c]{@{}l@{}}P: 0.65 \\ R: 0.39 \\ F1: 0.49\end{tabular} &  
\begin{tabular}[c]{@{}l@{}}P: 1.00 \\ R: 0.11 \\ F1: 0.20\end{tabular} &  
\begin{tabular}[c]{@{}l@{}}P: 0.00 \\ R: 0.00 \\ F1: 0.00\end{tabular} &  
\begin{tabular}[c]{@{}l@{}}P: 1.00 \\ R: 0.25 \\ F1: 0.40\end{tabular} &  
\begin{tabular}[c]{@{}l@{}}P: 1.00 \\ R: 0.25 \\ F1: 0.40\end{tabular} &  
\begin{tabular}[c]{@{}l@{}}P: 0.64 \\ R: 0.64 \\ F1: 0.59\end{tabular} \\ \midrule

Logistic Regression & 
\begin{tabular}[c]{@{}l@{}}P: 0.66 \\ R: 0.87 \\ F1: 0.75\end{tabular} &  
\begin{tabular}[c]{@{}l@{}}P: 0.66 \\ R: 0.52 \\ F1:\textbf{ 0.58}\end{tabular} &  
\begin{tabular}[c]{@{}l@{}}P: 1.00 \\ R: 0.11 \\ F1: 0.20\end{tabular} &  
\begin{tabular}[c]{@{}l@{}}P: 0.00 \\ R: 0.00 \\ F1: 0.00\end{tabular} &  
\begin{tabular}[c]{@{}l@{}}P: 0.00 \\ R: 0.00 \\ F1: 0.00\end{tabular} &  
\begin{tabular}[c]{@{}l@{}}P: 1.00 \\ R: 0.67 \\ F1:\textbf{
0.80}\end{tabular} &  
\begin{tabular}[c]{@{}l@{}}P: 0.66 \\ R: 0.66 \\ F1: 0.62\end{tabular} \\ \midrule

Naive Bayes & 
\begin{tabular}[c]{@{}l@{}}P: 0.67 \\ R: 0.79 \\ F1: 0.72\end{tabular} &  
\begin{tabular}[c]{@{}l@{}}P: 0.55 \\ R: 0.52 \\ F1: 0.53\end{tabular} &  
\begin{tabular}[c]{@{}l@{}}P: 1.00 \\ R: 0.22 \\ F1: 0.36\end{tabular} &  
\begin{tabular}[c]{@{}l@{}}P: 0.00 \\ R: 0.00 \\ F1: 0.00\end{tabular} &  
\begin{tabular}[c]{@{}l@{}}P: 0.67 \\ R: 0.50 \\ F1:\textbf{ 0.57}\end{tabular} &  
\begin{tabular}[c]{@{}l@{}}P: 1.00 \\ R: 0.67 \\ F1:\textbf{
0.80}\end{tabular} &  
\begin{tabular}[c]{@{}l@{}}P: 0.63 \\ R: 0.63 \\ F1: 0.61\end{tabular} \\ \midrule

KNN & 
\begin{tabular}[c]{@{}l@{}}P: 0.66 \\ R: 0.74 \\ F1: 0.70\end{tabular} &  
\begin{tabular}[c]{@{}l@{}}P: 0.61 \\ R: 0.55 \\ F1:\textbf{ 0.58}\end{tabular} &  
\begin{tabular}[c]{@{}l@{}}P: 1.00 \\ R: 0.22 \\ F1: 0.36\end{tabular} &  
\begin{tabular}[c]{@{}l@{}}P: 0.00 \\ R: 0.00 \\ F1: 0.00\end{tabular} &  
\begin{tabular}[c]{@{}l@{}}P: 1.00 \\ R: 0.25 \\ F1: 0.40\end{tabular} &  
\begin{tabular}[c]{@{}l@{}}P: 1.00 \\ R: 0.67 \\ F1:\textbf{
0.80}\end{tabular} &  
\begin{tabular}[c]{@{}l@{}}P: 0.64 \\ R: 0.63 \\ F1: 0.61\end{tabular} \\ \bottomrule
\end{tabular}}
\end{table*}

Table \ref{tab:metrics} presents the performance of seven ML models across various smell categories, evaluated by precision (P), recall (R), and F1-score. Each model's performance was assessed in the following categories: \textit{not a smell, obvious, task, beautification, vague, commented-out code} and \textit{Total} referring to the weighted average of all categories.

In the evaluation, Gradient Boosting and Random Forest models performed best in the \textit{not a smell} category, each achieving an F1-score of 0.77. Logistic Regression stood out in the \textit{obvious} category with the highest F1-score of 0.58. Gradient Boosting led in the \textit{task} category with an F1-score of 0.80. Although relatively low, the best score in the \textit{beautification} category was 0.57, recorded by Random Forest and Decision Tree. The remaining models demonstrated no detection capability in the \textit{beautification} category, each scoring 0.00. 

Three models, including Random Forest, Decision Tree, and Naive Bayes, achieved an F1 score of 0.57 detecting \textit{vague} smells. \textit{Commented-out code} smell was detected with an F1 score of 0.80 by the majority of the models, excluding Gradient Boosting and SVM.

Random Forest performed best, classifying the code comment smells according to precision, recall, and F1-score. Notably, the classifier achieved high recall indicating that it successfully detected the smell types.

\begin{table}[ht]
    \caption{Performance of ML Models Based on MCC and Accuracy}
    \centering
        \begin{tabular}{lcc}
        \hline
        \textbf{Model} & \textbf{MCC} & \textbf{Accuracy} \\
        \hline
        Gradient Boosting    & 0.38 & 0.66 \\
        Random Forest        & 0.44 & 0.69 \\
        Decision Tree        & 0.35 & 0.63 \\
        SVM                  & 0.30 & 0.64 \\
        Logistic Regression  & 0.32 & 0.65 \\
        Naive Bayes          & 0.30 & 0.63 \\
        KNN                  & 0.29 & 0.61 \\

        \hline
        \end{tabular}
    \label{tab:model_performance}
\end{table}

The results in Table \ref{tab:model_performance} show how well different ML models perform based on MCC and accuracy. Random Forest has the best overall performance, with an MCC of 0.44 and an accuracy of 0.69, making it the most reliable model in this study. Gradient Boosting also performs well, with an MCC of 0.38 and an accuracy of 0.66. Decision Tree, SVM, Logistic Regression, Naive Bayes, and KNN show moderate performance, with MCC values ranging from 0.29 to 0.35 and accuracy between 0.61 and 0.65. These models are fairly consistent, but not as strong as Random Forest or Gradient Boosting.

\subsection{Performance of GPT-4}

\textit{Using the Original Dataset:} We prompted the GPT-4 model to classify the inline code comments into ten distinct groups, which consisted of nine specific smell types (\textit{task, vague, beautification, irrelevant, commented-out code, non-local info, too much info, obvious, misleading}) and one non-smell type, labeled as \textit{not a smell}. Confusion matrix is presented in the replication package.\footnote{\url{https://figshare.com/s/49ae815ab21d916726ab?file=49678575}}

\begin{table}[ht]
    \caption{Performance of GPT-4 with Original Dataset}
    \centering
    \begin{tabular}{llllll}
        \toprule
        & Precision & Recall & F1-Score & Support \\
        \midrule
        Beautification & 0.57 & 0.67 & 0.62 & 49 \\
        Commented-out code & 0.20 & 0.03 & 0.05 & 35 \\
        Irrelevant & 0.11 & 0.75 & 0.19 & 4 \\
        Misleading & 0.12 & 0.06 & 0.08 & 16 \\
        Non-local info & 0.00 & 0.00 & 0.00 & 4 \\
        Not a smell & 0.65 & 0.42 & 0.51 & 1253 \\
        Obvious & 0.76 & 0.06 & 0.11 & 672 \\
        Task & 0.37 & 0.81 & 0.51 & 121 \\
        Too much info & 0.11 & 0.67 & 0.18 & 6 \\
        Vague & 0.05 & 0.78 & 0.09 & 51 \\
        \midrule
        Accuracy & & & 0.34 & 2211 \\
        Macro Avg & 0.29 & 0.43 & 0.23 & 2211 \\
        Weighted Avg & 0.64 & 0.34 & 0.37 & 2211 \\
        \bottomrule
    \end{tabular}
    \label{tab:performance_gpt4nocode}
\end{table}

The performance metrics for GPT-4 using the original dataset are presented in Table \ref{tab:performance_gpt4nocode}. The model achieved an overall accuracy of 0.34 across 2,211 instances. For the \textit{beautification} class, the precision was 0.57, indicating that 57\% of predicted instances were correctly identified, while the recall of 0.67 suggested that the model captured 67\% of actual instances. According to the confusion matrices, the \textit{commented-out code} class was predominantly predicted as \textit{not a smell}, resulting in a precision of 0.20 and a recall of 0.03. In the \textit{irrelevant} category, three out of four instances were correctly identified, while only one out of 16 instances of \textit{misleading} were accurately predicted. None of the \textit{non-local info} instances were predicted correctly. For \textit{not a smell} category, the model achieved a precision of 0.65 and a recall of 0.42, reflecting a moderate level of correct predictions but a limited ability to capture all relevant instances. The \textit{obvious} class exhibited high precision at 0.76 but a low recall of 0.06, indicating that while it effectively identified positive cases, it missed many actual instances. The \textit{task} class achieved a precision of 0.37 and a recall of 0.81, demonstrating a strong capability to recognize actual \textit{task} instances, although precision was lower. The \textit{too much info} instances were predicted with a recall of 0.67, indicating good identification of relevant cases. Lastly, the \textit{vague} class had a precision of 0.05 and a recall of 0.78, meaning that while the model identified a majority of the vague instances, it also misclassified numerous non-vague instances as vague.

\textit{Using the Augmented Dataset:} In this part of the experiment, given the augmented dataset, we prompted GPT-4 model to classify the inline code comments into ten categories: \textit{task, vague, beautification, irrelevant, not a smell, commented-out code, non-local info, too much info, obvious,} and \textit{misleading}. The confusion matrix is presented in the replication package.\footnote{\url{ https://figshare.com/s/49ae815ab21d916726ab?file=49678572}}

\begin{table}[ht]
    \caption{Performance of GPT-4 with the Augmented Dataset}
    \centering
    \begin{tabular}{llllll}
        \toprule
        & Precision & Recall & F1-Score & Support \\
        \midrule
        Beautification & 0.54 & 0.65 & 0.59 & 49 \\
        Commented-out code & 0.60 & 0.17 & 0.27 & 35 \\
        Irrelevant & 0.17 & 0.75 & 0.27 & 4 \\
        Misleading & 0.12 & 0.40 & 0.18 & 16 \\
        Non-local info & 0.00 & 0.00 & 0.00 & 4 \\
        Not a smell & 0.68 & 0.68 & 0.68 & 1253 \\
        Obvious & 0.55 & 0.30 & 0.39 & 672 \\
        Task & 0.48 & 0.71 & 0.57 & 121 \\
        Too much info & 0.21 & 0.50 & 0.30 & 6 \\
        Vague & 0.13 & 0.42 & 0.19 & 52 \\
        \midrule
        Accuracy & & & 0.55 & 2211 \\
        Macro Avg & 0.35 & 0.46 & 0.35 & 2211 \\
        Weighted Avg & 0.60 & 0.55 & 0.56 & 2211 \\
        \bottomrule
    \end{tabular}
    \label{tab:performance_gpt4withcode}
\end{table}

Table \ref{tab:performance_gpt4withcode} presents the performance of the GPT-4 model prompted with the augmented dataset, showing a notable improvement in overall accuracy from 34\% to 55\%. Additionally, the MCC score improved from 0.16 to 0.28, reflecting better overall predictive performance. For the \textit{beautification} class, the precision was 0.54, consistent with the performance of the model using the original dataset. This aligns with expectations, as \textit{task}, \textit{beautification}, and \textit{commented-out code} categories in the augmented dataset do not have a "code" section. Also, the \textit{task} class scored a precision of 0.48 and a recall of 0.71, which were similar to the previous scores. The \textit{commented-out code} class demonstrated a slight increase in correctly identified instances compared to the original dataset. 

In the \textit{irrelevant} category, the model achieved a precision of 0.17 and a recall of 0.75. The \textit{misleading} class showed improvements with a precision of 0.12 and a recall of 0.40, reflecting enhanced effectiveness in accurately identifying and capturing instances. 

The \textit{non-local info} class did not yield any correct predictions, resulting in a precision, recall, and F1-score of 0.00. For \textit{not a smell}, the model demonstrated a precision of 0.68 and a recall of 0.68, indicating significant improvement in identifying relevant instances. The model classified 853 instances correctly with the augmented dataset compared with the previous 527.
The \textit{obvious} class achieved a precision of 0.55 and a recall of 0.30, successfully predicting 207 instances compared to just 41 in the previous results. 

The \textit{too much info} class had a precision of 0.21 and a recall of 0.50, showing limited improvement. Lastly, the \textit{vague} class recorded a precision of 0.13 and a recall of 0.42, capturing some vague instances but misclassifying 154 non-vague instances as vague—a reduction from the previous misclassification of 779 instances. Previously, 430 \textit{not a smell} instances were classified as \textit{vague}; after the augmentation, this dropped to 67. Similarly, 322 \textit{obvious} smells were predicted as \textit{vague}, but when the code context was given, this number decreased to 69.

The weighted averages indicated a precision of 0.60, a recall of 0.55, and an F1-score of 0.56, demonstrating overall effectiveness while accounting for the distribution of instances in each class.

%The support column in the table indicates the number of instances for each comment category within the augmented dataset, which influences the model's learning and classification accuracy.

\section{Discussion}\label{sec:section5}

%In this section, we discuss the performance of ML models and the GPT-4 model.
%In-depth discussion of research findings.
%Comparison with existing literature and related work.
%Implications and significance of the results.
%Discussion of any limitations encountered in the study.
% \subsection{Heuristics}
% The heuristics for detecting \textit{task} smells exhibit a notable F1-score of 0.68, outperforming six out of the seven machine learning models evaluated for this specific comment smell type. However, Gradient Boosting emerged as the top performer with an F1-score of 0.74. When compared to the GPT-4 model, which achieved F1-scores of 0.58 and 0.61 on different datasets for \textit{task} smell detection, the heuristics demonstrated a slight performance advantage. Developers may find the heuristics valuable for detecting \textit{task} smells.

% The heuristics applied for detecting the \textit{beautification}  comment smell type achieved an F1-score of 0.40, demonstrating its superiority over machine learning models. Among these models, Naive Bayes performed the best for detecting *beautification* comments, achieving an F1-score of 0.22. Furthermore, the heuristics outperformed the GPT-4 model on both the original and augmented datasets, which attained F1-scores of 0.03 and 0.13, respectively, for this comment smell type. These results suggest that heuristics can effectively identify the \textit{beautification}  comment smell type.

\subsection{Performance of ML Models}
These findings provide insights into the comparative effectiveness of different models for inline code comment smell detection.

\textbf{Gradient Boosting:} This model demonstrates strong performance, particularly in the \textit{not a smell} and \textit{task} categories, achieving a precision of 0.71 and 1.00, respectively. This indicates it can differentiate non-problematic code and identify tasks well. However, it struggles with \textit{beautification}, where it has no true positives.

\textbf{Random Forest:} Random Forest shows high precision and recall across some categories, especially excelling in \textit{not a smell} (P: 0.71, R: 0.85) and \textit{commented-out code} (P: 0.40, R: 1.00). It achieves the highest overall F1-score of 0.67.

\textbf{Decision Tree:} With balanced performance, it achieves a precision of 0.74 in the \textit{not a smell} category, but its effectiveness drops in \textit{beautification} and \textit{vague}.

\textbf{Logistic Regression and SVM: }  Both models exhibit moderate performance, with Logistic Regression achieving a precision of 0.66 in \textit{not a smell} but faltering in more complex categories. SVM shows similar trends, with a precision of 0.63 for \textit{not a smell} but low scores for others.

\textbf{Naive Bayes and KNN:} These models rank lower, with Naive Bayes having a precision of 0.67 for \textit{not a smell} but lower performance in other categories. KNN faces similar challenges, performing best in \textit{obvious} smells but struggling overall.

Overall, Random Forest and Gradient Boosting are the most reliable models for detecting inline code comment smells, while Naive Bayes, SVM, and KNN show limitations in capturing complex patterns. Random Forest's strong performance across multiple categories can be attributed to its ensemble nature, which allows it to combine the predictions of multiple decision trees and reduce overfitting. It handles noisy data and high-dimensional features well, making it effective for this multi-class problem.

Manual pattern analysis in the ML classifiers was not feasible due to the use of techniques like 10-fold cross-validation, SMOTE oversampling, and train-test-validation splits. These methods shuffle and manipulate the dataset, making it difficult to track specific instances throughout the prediction process. As a result, individual patterns or behaviors within the dataset become less traceable, preventing direct manual inspection of how specific data points influence model performance.

% \[
% \text{Random‑baseline accuracy} = \frac{100}{\lvert C \rvert} = 10\% \qquad (\lvert C \rvert = 10)
% \]

% Comparing the 10\% random baseline with the classical models in Table \ref{tab:model_performance} shows that every conventional algorithm achieves at least a 6x lift (60\% accuracy), demonstrating that—even though these techniques are several years old—they still extract meaningful signal from our dataset.

\subsection{Performance of GPT-4}
%Since our dataset is imbalanced, recall is more crucial than precision to avoid missing minority class instances. In detecting code comment smells, missing problematic comments can lead to bugs or maintenance issues. Prioritizing recall ensures the model catches as many issues as possible, even at the expense of some false positives. This is vital in software development to mitigate risks and maintain code quality.

Through manual analysis, several important findings regarding GPT-4’s performance in predicting code comment smells have emerged. One key observation is the significant improvement in the accuracy of predicting comments labeled as \textit{not a smell}. In the original dataset, comments labeled as \textit{not a smell} were predicted accurately 527 times. Notably, in the augmented dataset, this number increased to 853 correct predictions. This indicates that when code is provided, GPT-4 demonstrates a better understanding of whether a comment is part of a smell or not. Without the code, the model struggled to determine whether a comment was a smell.
Similarly, another important finding is the improvement in the detection of \textit{obvious} smells. The results show that with the original dataset, \textit{obvious} comments were predicted correctly 41 times, while with the augmented dataset, this number increased to 207. This significant increase may be attributed to the model's ability to better understand the comment’s context when code is present, making it easier to determine if a comment is \textit{obvious}.
Another major contribution of the augmented dataset is the reduction in misclassification of \textit{not a smell} comments as \textit{vague} smells. In the original dataset, \textit{not a smell} was incorrectly predicted as \textit{vague} 430 times, but in the augmented dataset, this number dropped to 67. This suggests that without context, GPT-4 tends to label unclear comments as \textit{vague}, as it cannot fully grasp their purpose or clarity. With the code, the model can more accurately assess whether a comment is truly \textit{vague}.
Additionally, the number of \textit{obvious} comments misclassified as \textit{vague} dramatically decreased. In the original dataset, this occurred 322 times, but in the augmented dataset, the number dropped to 69. This shows that when code context is missing, GPT-4 tends to label comments as \textit{vague}, but when code is introduced, it recognizes that these comments are not actually \textit{vague}.

The overall impact of the augmented dataset is also evident in the recall across most classes, including \textit{misleading, not a smell} and \textit{obvious}. High recall is crucial, especially given the dataset's inherent imbalance, as it ensures the model captures the most problematic comments and minimizes the risk of overlooking critical issues.

Furthermore, for the \textit{beautification} smell, it is important to note that since we do not provide code snippets in the augmented dataset for this category, there is no difference between the original and augmented datasets for these comments. The decrease of 0.03 in the F1 score for \textit{beautification} comments can be attributed to the stochasticity of GPT-4, rather than any significant change in model understanding.

\begin{table}[ht]
\caption{Increase in F1-scores of GPT-4 model using the augmented dataset}
\centering
\begin{tabular}{ll}
\hline
\textbf{Category} & \textbf{Increase in F1-Score} \\
\hline
Beautification & -0.03 \\
Commented-out code & 0.22 \\
Not a smell & 0.17 \\
Obvious & 0.28 \\
Task & 0.06 \\
Vague & 0.10 \\
Misleading & 0.10 \\
Too much info & 0.12 \\
Non-local info & 0.00 \\
Irrelevant & 0.08 \\
\hline
\end{tabular}
\label{tab:improvement}
\end{table}

Furthermore, Table \ref{tab:improvement} presents the improvements in F1-scores of GPT-4 model predictions comparing the augmented dataset with the original one \cite{elgun2024}. Eight out of ten categories improved. In contrast, the \textit{non-local info} category did not improve, maintaining a zero increase, which might indicate a limitation in the model's capability to interpret the original places of these comments. \textit{Not a smell} category improved by 17\%, implying that the code segments given in the augmented dataset help distinguish non-problematic comments from actual comment smells. 

A dramatic improvement was observed in the F1-score for \textit{obvious} smells, which improved from 11\% to 39\%. This suggests that when the model processes both code and comments together, it can more effectively detect discrepancies between them. The augmented dataset effectively assists the model in identifying code that could potentially mislead developers or maintainers. Conversely, categories such as \textit{beautification, task} and \textit{commented-out code} showed only slight differences from the original dataset, as they were not augmented with related code context. \textit{Vague, misleading, too much info}, and \textit{irrelevant} categories have modest improvements in F1-score when detecting the corresponding smells, which suggests that the augmented dataset helps the model slightly better understand or classify code related to formatting and stylistic enhancements. 

\sloppy
While the augmented dataset improved many aspects of performance, analysis through the confusion matrix revealed ongoing issues. For instance, the model only correctly predicted \textit{commented-out code} smells for Java comments. We believe this limitation arises from the nature of our prompt, where we described the \textit{commented-out code} smell with a specific Java example: \textit{A code piece that is commented out}, accompanied by the example \textit{//facade.registerProxy(newSoundAssetProxy());}. This Java-centric example may explain why the model struggled to identify the commented-out code smell in Python comments.

%Another problem is that the model mispredicted \textit{obvious} smell as \textit{not a smell} 245 times with the original dataset, but, contrary to our expectation, with the augmented dataset, this number increased to 352. The root cause of this problem can be the definition of the \textit{not a smell} category. In the prompt, the \textit{not a smell} category is described as:
%\textit{
%\begin{itemize}
 %   \item Provides useful context or explanation that enhances code comprehension.
  %  \item Clarifies the purpose or intent of the code.
  %  \item Warns about potential issues or edge cases.
   % \item Documents complex algorithms or business logic.
    %\item Follows best practices for commenting style and clarity.
%\end{itemize}
%}
%The \textit{obvious} smell type does not contradict the definition of the \textit{not a smell}. This may be an indication of why \textit{obvious} smell type is misclassified as \textit{not a smell}.

Another issue is that GPT-4 failed to detect the \textit{non-local information} smell. This may be because identifying \textit{non-local information} requires access to the entire file, as the smell involves comments referencing parts of the code that are not nearby. However, in the augmented dataset, the associated code segment provided was only where the comment was written, which may have limited GPT-4's ability to identify the smell correctly. Additionally, the small number of instances (only four) of \textit{non-local information} in the dataset may have contributed to the model's difficulty in learning this pattern effectively.

%Reflecting on these findings, we believe that our prompts and examples could be improved. The current effectiveness seems constrained by our strict adherence to the taxonomy, which may limit the model’s ability to handle different types of code comment smells more flexibly.

\begin{table}[ht]
    \caption{Results of Best Performing ML Model \& GPT-4}
    \centering
    \resizebox{\columnwidth}{!}{%
        \begin{tabular}{lccccc}
        \hline
        \textbf{Model} & \textbf{MCC} & \textbf{Accuracy} & \textbf{Precision} & \textbf{Recall} & \textbf{F1} \\
        \hline
        Random Forest & 0.44 & 0.69 & 0.70 & 0.69 & 0.67 \\
        \hline
        GPT-4 with Augmented Data & 0.28 & 0.55 & 0.61 & 0.55 & 0.56 \\
        \hline
        \end{tabular}%
    }
    \label{tab:final_comparison}
\end{table}

Table \ref{tab:final_comparison} presents the performances of the best-performing ML model, Random Forest, with GPT-4 on the augmented dataset in terms of MCC, accuracy, precision, recall, and F1. One possible reason for Random Forest's performance is its ability to handle structured data effectively, which aligns well with the features in the dataset. GPT-4 may struggle with smaller, domain-specific datasets where it cannot leverage its full potential when the data or the prompt lacks enough diversity or examples for the model to generalize effectively. It is important to note that GPT-4 makes predictions across ten categories, while the ML models predict across six categories due to dataset filtering, which involves removing minority classes.

\subsection{Implications for Practitioners and Future Research}

This study presents several potential implications for software development practitioners, focusing on the integration and application of automated tools for enhancing code documentation quality:

\textbf{Enhanced Detection Tools}: By incorporating the ML models and GPT-4 model demonstrated in this study, practitioners can potentially reduce the incidence of code comment smells. These models can be integrated into existing automated systems within development environments to streamline the maintenance of high-quality code documentation. However, it is important to note that the current accuracy levels of our models are not as high as one would ideally desire, underscoring the need for further refinement and experimentation. Such tools could be integrated into the development workflows in various ways, as follows:

\begin{itemize}[nosep, leftmargin=*]
    \item Continuous Integration/Continuous Deployment (CI/CD) Systems: Integrating our models into CI/CD pipelines allows for real-time scanning and updating of code comments as code evolves, ensuring documentation accuracy throughout the development process.
    \item Code Review Processes: By implementing these models within code review tools, potential comment smells can be automatically flagged before merging, aiding reviewers in maintaining consistent and high-quality documentation.
    \item IDE Plugins: Development of IDE plugins that utilize our detection models can provide immediate feedback to developers, suggesting corrections or improvements to comments directly within the development environment.
\end{itemize}

\textbf{Maintainability and Technical Debt Reduction:} Accurate and up-to-date code comments directly contribute to software maintainability and prevent the accumulation of technical debt associated with outdated documentation.

\textbf{Enhanced Developer Productivity:} Automated tools reduce the time developers spend deciphering outdated comments, allowing them to focus on core development tasks and enhancing overall productivity and efficiency.

\textbf{Future Research:} The augmented dataset, which has significantly improved the performance of large language models (GPT-4—from 34\% to 55\% in accuracy) demonstrates the advantages of enriching datasets with comprehensive code-comment pairs. This approach is essential for training more effective models and can be a pivotal strategy for continuous improvement, ensuring that documentation evolves in tandem with code development. The augmented dataset and associated code artifacts are made available online, providing practitioners with valuable resources to implement or benchmark their automated code comment smell detection tools.

%\textbf{Case Studies and Practical Examples:} We plan to include case studies detailing successful implementations of our techniques in various organizational contexts. These will provide a practical roadmap for teams interested in similar integrations, illustrating the setup, integration, and positive impacts on project outcomes.

\section{Threats to Validity}\label{sec:section6}
%This section discusses the potential threats to the validity of our study by addressing issues related to internal and external validity.

\subsection{Internal Validity}

\textit{Taxonomy and Dataset Dependence:} In alignment with Jabrayilzade et al.\cite{elgun2024}, it is important to acknowledge that the detection of code comment smells is inherently dependent on the classification of smell types. Variations in taxonomy could impact the identification process.
The adopted dataset benefited from manual labeling by three individuals, recognizing the potential for bias and limited perspectives is essential. Moreover, the taxonomy assumes that each inline comment belongs to one smell category. We chose to classify each comment into a single smell type, which aligns with the labeling provided in the adopted dataset. This approach helps maintain consistency in our analysis, but may not fully reflect the complexity of real-world coding practices, where comments might exhibit multiple smells simultaneously.

\textit{Bias of Models:} Imbalanced distribution of code comment smells in our dataset created challenges, particularly for the \textit{beautification} smell type, which constituted only 2.2\% of the dataset. Five out of the seven ML models consistently scored 0\% in accuracy, precision, and F1 for this category, demonstrating the impact of data distribution on model performance and the challenges in identifying underrepresented classes.

\textit{Improvement of the Dataset:} To improve prediction precision, we added related code segments to the comments, but manual selection risks inaccuracies in code-comment pairing, requiring cautious interpretation of results.

\textit{OpenAI GPT-4 Model:} We compared GPT-4 model predictions to demonstrate the benefits of data augmentation. Even with careful prompt engineering and parameter tuning, large language models face limitations such as an overreliance on surface patterns, a lack of interpretability, and constrained domain-specific knowledge\cite{hadi_tashi_qureshi_shah_muneer_irfan_zafar_shaikh_akhtar_wu_etal._2023}. %One another potential threat to validity is that our GPT-4 prompts were strictly based on the examples and definitions from Jabrayilzade et al.'s taxonomy \cite{elgun2024}, without any modifications, including no prompt engineering. This may limit the model's ability to detect variations or nuances in code comment smells that fall outside of the original examples, potentially reducing the generalizability of our results.

% OpenAI GPT-4 Model: We compared the results of the OpenAI
% GPT-4 model’s predictions to show the enhancements gained by
% data augmentation. Although we utilized prompt engineering skills
% and fine-tuning, using LLM’s introduces several limitations, includ-
% ing overreliance on surface-level patterns, lack of interpretability,
% and limited domain-specific knowledge [3].

\subsection{External Validity}
 
The generalizability of our findings may be influenced by the characteristics and nature of the selected dataset \cite{elgun2024}.  The size and composition of the dataset may not fully capture the diversity of coding practices in larger software ecosystems since eight projects were examined. This could affect the broader relevance of the identified code comment smells. Additionally, the dataset featured projects coded in Java and Python. The exclusion of other languages may introduce bias and limit the broader applicability of our identified code comment smells.

The size of our dataset, while carefully labeled and insightful, poses an inherent limitation to the external validity of our study. The finite scale of the dataset may restrict the generalizability of our findings and limit the robustness of the identified code comment smells. A larger dataset could potentially offer a more comprehensive exploration of code comment smells, capturing a broader array of patterns and scenarios in software development. 

The exclusive use of GPT-4 in our comparison, as its specific architecture and training data may influence the outcomes and limit the generalizability of our findings to other large language models or traditional machine learning approaches.

% The insights derived from a more expansive dataset would enhance the scope of our study, providing a stronger foundation for drawing meaningful conclusions about the prevalence and characteristics of inline code comment smells.

\section{Conclusion and Future Work}\label{sec:section7}
%Summary of the key points and findings
%Contribution of the research to the field
%Practical implications of the study
%Concluding remarks and insights
%Directions for future research and areas of further investigation
In this study, we enhanced the automated detection of inline code comment smells by utilizing ML models and a large language model. Building on the taxonomy and manually labeled dataset provided by previous research, we augmented the dataset by pairing comments with code segments to improve the code comment smell detection.

We prompted the GPT-4 model with both the original and augmented dataset and observed a 21\% increase in the overall accuracy of smell type classification.  With the original dataset, GPT-4 struggled to identify correlations between different smell types and the given comments, which limited its accuracy to just 34\%. When using the augmented dataset, GPT-4 successfully labeled 55\% of the comments correctly across ten categories, nine of which are smell categories and a \textit{not a smell} category. MCC score increased from 0.16 to 0.28.
%We adapted our model training approach by excluding minority classes to address the imbalanced dataset, which reduced the number of smell types considered from nine to five. Despite this adaptation, the detection of \textit{beautification} and \textit{vague} smells underperformed, likely due to the dataset's limited size and the residual imbalance.
Our work marks a significant step in code comment smell detection, establishing a baseline with models like Random Forest, Gradient Boosting, and Logistic Regression, achieving accuracies of 69\%, 66\%, and 65\%, respectively, and Random Forest achieving an MCC score of 0.44.

%Our future work involves developing tools to mitigate code comment smells, such as an IDE extension for real-time detection and suggestions and a GitHub bot that analyzes pull requests for potential comment smells. These initiatives aim to enhance code comment smell detection, improving developer practices and code quality.

% Our future work will focus on developing practical tools to mitigate code comment smells. We plan to introduce a VSCode extension that offers real-time detection and suggestions for developers, enhancing their ability to identify and address smells across entire projects. Additionally, we propose the creation of a GitHub bot to analyze pull requests for code comment smells, equipped with a notification system to alert developers of potential issues.

% These future endeavors underscore our commitment to advancing code comment smell detection, tackling practical challenges in software engineering, and delivering valuable tools that improve developer practices and code quality.

\section{Data Availability}  \label{dataAvailability}
All datasets and code used in this study are available at \url{https://figshare.com/s/49ae815ab21d916726ab}.
%\\

\newpage

\bibliographystyle{ACM-Reference-Format}
\bibliography{references}

\end{document}